\documentclass[showpacs,preprintnumbers,amsmath,twocolumn,amssymb]{revtex4}
\usepackage{graphicx}
\usepackage{dcolumn}
\usepackage{bm}
\usepackage{amssymb}
\usepackage{array}
\usepackage{hhline}
\usepackage{longtable}

\begin{document}

\title{Exact analytical solution of average path length for Apollonian networks}

\author{Zhongzhi Zhang$^{1,2}$}

\author{Lichao Chen $^{1,2}$}

\author{Shuigeng Zhou$^{1,2}$}
\email{sgzhou@fudan.edu.cn}

\author{Lujun Fang$^{1,2}$}

\author{Jihong Guan$^{3}$}

\author{Tao Zou$^{1,2}$}

\affiliation {$^{1}$Department of Computer Science and Engineering,
Fudan University, Shanghai 200433, China}

\affiliation {$^{2}$Shanghai Key Lab of Intelligent Information
Processing, Fudan University, Shanghai
200433, China} %

\affiliation{$^{3}$Department of Computer Science and Technology,
Tongji University, 4800 Cao'an Road, Shanghai 201804, China}


\begin{abstract}
With the help of recursion relations derived from the
self-similar structure, we obtain the exact solution of average path
length, $\bar{d}_t$, for Apollonian networks. In contrast to the
well-known numerical result $\bar{d}_t \propto (\ln N_t)^{3/4}$
[Phys. Rev. Lett. \textbf{94}, 018702 (2005)], our rigorous solution
shows that the average path length grows logarithmically as
$\bar{d}_t \propto \ln N_t$ in the infinite limit of network size
$N_t$. The extensive numerical calculations  completely agree with
our closed-form solution.

\end{abstract}

\pacs{89.75.Hc, 89.75.Da, 02.10.Ox, 05.10.-a}
\date{\today}
\maketitle

One of the most important properties of complex networks is average
path length (APL), which is the mean length of the shortest paths
between all pairs of vertices (nodes)~\cite{CoRoTrVi07}. Most real
networks have been shown to be small-world or ultra small-world
networks~\cite{AlBa02,DoMe02,Ne03,BoLaMoChHw06}, that is, their APL
$d$ behaves a logarithmic or double logarithmic scaling with the
network size $N$: $d\sim \ln N$~\cite{WaSt98} or $d\sim \ln \ln
N$~\cite{CoHa03}. It has been established that APL is relevant in
many fields regarding real-life networks. In the design or
interpretation of routes in architectural design, signal integrity
in communication networks, the propagation of diseases or beliefs in
social networks or of technology in industrial networks, APL is a
natural network statistic to compute and interpret. It is strongly
believed that many processes such as routing, searching, and
spreading become more efficient when APL is smaller. So far, much
attention has been paid to the question of
APL~\cite{ChLu02,DoMeSa03,FrFrHo04,Lolo03,HoSiFrFrSu05,DoMeOl06}.

Recently, on the basis of the well-known Apollonian
packing~\cite{bo73}, Andrade \emph{et al.} introduced Apollonian
networks~\cite{AnHeAnSi05} which were also proposed by Doye and
Massen in Ref.~\cite{DoMa05} simultaneously. Apollonian networks
belong to a deterministic growing type of networks, which have drawn
much attention from the scientific communities and have turned out
to be a useful
tool~\cite{BaRaVi01,DoGoMe02,JuKiKa02,CoFeRa04,ZhRoZh07,RaSoMoOlBa02,RaBa03,CoOzPe00,ZhRoGo06,BeOs79,Hi07,ZhZhZo07,BaFeDa06,ZhZhFaGuZh07,RoHaAv07}.
 Many topological properties of Apollonian networks such as degree
distribution, clustering coefficient, and correlations have been
determined analytically~\cite{AnHeAnSi05,DoMa05}, and the effects of
the Apollonian networks on several dynamical models have been
intensively studied, including Ising model and a magnetic
model~\cite{AnHeAnSi05,AnHe05,LiGaHe04,ZhRoZh06}. Despite the
importance and usefulness of the quantity APL, there is no
analytical calculations for the APL of Apollonian networks.

In this report, we derive an exact formula for the average path
length characterizing the Apollonian networks. The analytic method
is based on the recursive construction and self-similar structure of
Apollonian networks. Our rigorous result shows that APL grows
logarithmically with the number of nodes. The obtained analytical
solution modifies the previous numerical result
in~\cite{AnHeAnSi05}, where the authors claimed that the APL of
Apollonian networks scales sub-logarithmically with network size.
Our analytical technique could provide a paradigm for computing the
APL of deterministic networks.

\begin{figure}
\begin{center}
\includegraphics[width=8cm]{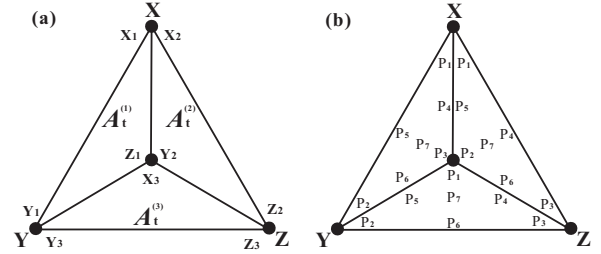}
\caption{(a) Second construction method of Apollonian network.
$A_{t+1}$, is obtained by joining three copies of
    $A_t$ denoted as
    $A_t^{(\varphi)}$ $(\varphi=1,2,3)$, which are
    connected to one another at the edge nodes (i.e., $X$, $Y$, $Z$, and $O$). Note that the central
node is represented as $O$ (marker not visible). (b) Illustration of
the recursive definition of node classification. From the node
classification of $A_t^{(1)}$, $A_t^{(2)}$, and $A_t^{(3)}$, we can
derive recursively the classification of nodes in network
$A_{t+1}$.} \label{copy}
\end{center}
\end{figure}

The Apollonian network, denoted as $A_t$ ($t\geq 0$) after $t$
generations, is constructed as follows~\cite{AnHeAnSi05}: For $t=0$,
$A_t$ is a triangle. For $t\geq 1$, $A_t$ is obtained from
$A_{t-1}$. For each of existing triangles created at step $t-1$, we
add a new vertex and join it to all the vertices of this triangle.
Alternatively, Apollonian network can be also created in another
method~\cite{Bobe05,HiBe06}. Given the generation $t$, $A_{t+1}$ may
be obtained by joining at four edge nodes three copies of $A_{t}$,
see Fig.~\ref{copy}(a). According to the latter construction
algorithm, we can easily compute the total number of vertices of
$A_t$ is $N_{t}=(3^{t}+5)/2$.

Apollonian network presents the typical characteristics of real-life
networks in nature and society~\cite{WaSt98,BaAl99}. It has a
power-law degree distribution with exponent $\gamma=1+\ln3/\ln
2$~\cite{AnHeAnSi05,DoMa05}, which belongs to the interval between 2
and 3. For any individual vertex with degree $k$, its clustering
coefficient $C(k)$ is also approximately inversely proportional to
its degree $k$ as $C(k)=(6/k)-[2/(k-1)]$. The mean value $C$ of
clustering coefficients of all vertices is very large, which
asymptotically reaches a constant value 0.8284. Moreover, the
network is small world, and its diameter, defined as the longest
shortest path length between all pairs of vertices, increases
logarithmically with the number of vertices~\cite{ZhCoFeRo06}. In
fact, Apollonian network can be expanded to general
cases~\cite{ZhCoFeRo06,DoMa05,ZhRoZh06} associated with other
self-similar packings~\cite{MaHeRi04}. Analogously, we can construct
the stochastic version of these
networks~\cite{ZhRoZh06,ZhYaWa05,ZhRoCo06,ZhZh07}. The main
topological properties of the Apollonian type networks are
controlled by the dimension of packings.

After introducing the Apollonian networks, we now investigate
analytically the average path length. We represent all the shortest
path lengths of network $A_{t}$ as a matrix in which the entry
$g_{uv}$ is the shortest path length from node $u$ to $v$. A measure
of the typical separation between two nodes in $A_{t}$ is given by
the average path length $\bar{d}_{t}$ defined as the mean of
shortest path lengths over all couples of nodes, i.e., $\bar{d}_{t}=
\frac{D_t}{N_t(N_t-1)/2}$, where $D_t = \sum_{u,v \in A_{t}, u \neq
v} g_{uv}$ denotes the sum of the shortest path length between two
nodes over all pairs.

By the second construction, it is obvious that Apollonian network
has a self-similar structure. Thus, the total distance $D_{t+1}$
satisfies the recursion relation
\begin{equation}\label{total02}
  D_{t+1} = 3\, D_t + \Theta_t-3,
\end{equation}
where $\Theta_t$ is the sum over all shortest path length whose
endpoints are not in the same $A_t$ branch. The solution of
Eq.~(\ref{total02}) is
\begin{equation}\label{total03}
  D_t = 3^{t-1}\, D_1 + \sum_{m=1}^{t-1} \left(3^{t-m-1} \Theta_m\right)-3^{t-1}.
\end{equation}
Thus, all that is left to obtain $D_t$ is to compute $\Theta_m$.

The paths that contribute to $\Theta_t$ must all go through at least
one of the four edge nodes ($X$, $Y$, $Z$, and $O$) at which the
different $A_t$ branches ($A_{t}^{(1)}$, $A_{t}^{(2)}$,
$A_{t}^{(3)}$) are connected. The analytical expression for
$\Theta_t$, named the crossing path length, can be derived as below.

Denote $\Theta_t^{\alpha,\beta}$ as the sum of all shortest paths
with endpoints in $A_t^{(\alpha)}$ and $A_t^{(\beta)}$. Note that
$\Theta_t^{\alpha,\beta}$ rules out the paths with endpoint at the
edge of $A_t^{(\alpha)}$ and $A_t^{(\beta)}$. For example, each path
contributed to $\Theta_t^{1,2}$ should not end at node $O$ or $X$.
Then the total sum $\Theta_t$ is given by $\Theta_t =\Theta_t^{1,2}
+ \Theta_t^{1,3} + \Theta_t^{2,3}$. By symmetry, $\Theta_t^{1,2} =
\Theta_t^{1,3} = \Theta_t^{2,3}$, so that $\Theta_t = 3\,
\Theta_t^{1,2}$, where $\Theta_t^{1,2}$ is given by the sum

\begin{equation}\label{cross03}
  \Theta_t^{1,2} = \sum_{\substack{u \in A_t^{(1)},\,\,v\in
      A_t^{(2)},\, u,v \ne X \bigcup u,v \ne O}} g_{uv}.
\end{equation}

To calculate the crossing path length $\Theta_t^{1,2}$, we classify
nodes in network $A_{t+1}$ into seven different parts according to
their shortest path lengths to each of the three vertices (i.e. $X$,
$Y$, $Z$) of the peripheral triangle $\triangle XYZ$. Notice that
vertices $X$, $Y$, and $Z$ themselves are not classified into any of
the seven parts represented as $P_{1}$, $P_{2}$, $P_{3}$, $P_{4}$,
$P_{5}$, $P_{6}$, and $P_{7}$, respectively. The classification of
nodes is shown in Fig.~\ref{copy}(b). For any interior node $v$, we
denote the shortest path lengths from $v$ to $X$, $Y$, $Z$ as $a$,
$b$, and $c$, respectively. By construction, $a$, $b$, $c$ can
differ by at most $1$ since vertices $X$, $Y$ and $Z$ are adjacent.
Then the classification function $class(v)$ of a node $v$ is defined
to be
\begin{equation}\label{classification}
class(v)=\left\{
\begin{array}{lc}
{\displaystyle{P_{1}}}
& \quad \hbox{for}\ a<b=c,\\
{\displaystyle{P_{2}}}
& \quad \hbox{for}\ b<a=c,\\
{\displaystyle{P_{3}}}
& \quad \hbox{for}\ c<a=b,\\
{\displaystyle{P_{4}}}
& \quad \hbox{for}\ a=c<b,\\
{\displaystyle{P_{5}}}
& \quad \hbox{for}\ a=b<c,\\
{\displaystyle{P_{6}}}
& \quad \hbox{for}\ b=c<a,\\
{\displaystyle{P_{7}}}
& \quad \hbox{for}\ a=b=c.\\
\end{array} \right.
\end{equation}
It should be mentioned that the definition of node classification is
recursive. For instance, Class $P_{1}$ and $P_{4}$ in $A_t^{(1)}$
belongs to Class $P_{1}$ in $A_{t+1}$, Class $P_{2}$ and $P_{6}$ in
$A_t^{(1)}$ belongs to Class $P_{2}$ in $A_{t+1}$, Class $P_{3}$ in
$A_t^{(1)}$ belongs to Class $P_{7}$ in $A_{t+1}$, and Class $P_{5}$
and $P_{7}$ in $A_t^{(1)}$ belongs to Class $P_{5}$ in $A_{t+1}$.
Since the three vertices $X$, $Y$ and $Z$ are symmetrical, in
Apollonian network we have the following equivalent relations from
the viewpoint of class cardinality: Classes $P_{1}$, $P_{2}$, and
$P_{3}$ are equivalent to one another, and it is the same with
Classes $P_{4}$, $P_{5}$, and $P_{6}$. We denote the number of nodes
in network $A_{t}$ that belong to Class $P_{1}$ as $N_{t,P_{1}}$,
the number of nodes in Class $P_{2}$ as $N_{t,P_{2}}$, and so on. By
symmetry, we have $N_{t,P_{1}}=N_{t,P_{2}}=N_{t,P_{3}}$ and
$N_{t,P_{4}}=N_{t,P_{5}}=N_{t,P_{6}}$. Therefore in the following
computation we will only consider $N_{t,P_{1}}$, $N_{t,P_{4}}$, and
$N_{t,P_{7}}$. It is easy to conclude that
\begin{eqnarray}
N_{t} &=& N_{t,P_{1}}+N_{t,P_{2}}+N_{t,P_{3}}+N_{t,P_{4}}
\nonumber\\ &\quad& \quad\quad\quad+N_{t,P_{5}}+N_{t,P_{6}}+N_{t,P_{7}}+3\nonumber\\
 &=&3\,N_{t,P_{1}} + 3\,N_{t,P_{4}} + N_{t,P_{7}}+3.
\end{eqnarray}
Considering the self-similar structure of Apollonian network, we can
easily know that at time $t+1$, the quantities $N_{t+1,P_{1}}$,
$N_{t+1,P_{4}}$, and $N_{t+1,P_{7}}$ evolve according to the
following recursive equations
\begin{eqnarray}\label{Np01}
\left\{
\begin{array}{ccc}
N_{t+1,P_{1}} &=& 2\,N_{t,P_{1}} + 2\,N_{t,P_{4}}\,, \\
N_{t+1,P_{4}} &=& N_{t,P_{4}} + N_{t,P_{7}}\,, \\
N_{t+1,P_{7}} &=& 3\,N_{t,P_{1}} + 1\,, \\
 \end{array}
 \right.
\end{eqnarray}
where we have used the equivalence relations:
$N_{t,P_{1}}=N_{t,P_{2}}=N_{t,P_{3}}$ and
$N_{t,P_{4}}=N_{t,P_{5}}=N_{t,P_{6}}$.

For a node $v$ in network $A_{t+1}$, we are also interested in the
smallest value of the shortest path length from $v$ to any of the
three peripheral vertices $X$, $Y$, and $Z$. We denote the shortest
distance as $f_v$, which can be defined to be $f_v = min(a,b,c)$.
Let $d_{t,P_{1}}$ denote the sum of $f_v$ of all nodes belonging to
Class $P_{1}$ in network $A_{t}$. Analogously, we can also define
the quantities $d_{t,P_{2}}$, $d_{t,P_{3}}$, $\cdots$,
$d_{t,P_{7}}$. Again by symmetry, we have
$d_{t,P_{1}}=d_{t,P_{2}}=d_{t,P_{3}}$,
$d_{t,P_{4}}=d_{t,P_{5}}=d_{t,P_{6}}$, and $d_{t,P_{1}}$,
$d_{t,P_{4}}$, $d_{t,P_{7}}$ can be written recursively as follows:
\begin{eqnarray}\label{dp01}
\left\{
\begin{array}{ccc}
d_{t+1,P_{1}} &=& 2\,d_{t,P_{1}} +2\, d_{t,P_{4}}\,, \\
d_{t+1,P_{4}} &=& d_{t,P_{4}} + d_{t,P_{7}}\,, \\
d_{t+1,P_{7}} &=& 3\,(d_{t,P_{1}} + N_{t,P_{1}}) + 1\,. \\
 \end{array}
 \right.
\end{eqnarray}
Next we begin to determine the value of the crossing path length
$\Theta_t^{1,2}$.  Using the definition of $f_v$, $\Theta_t^{1,2}$
can be rewritten as
\begin{equation}\label{cross04}
  \Theta_t^{1,2} = \sum_{\substack{u \in A_t^{(1)},\,\,v\in
      A_t^{(2)},\, u,v \ne X \bigcup u,v \ne O}} (f_u+f_v+\delta_{uv})\,,
\end{equation}
where $\delta_{uv}$ can be 0 or 1 depending on whether the two paths
corresponding to $f_u$ and $f_v$ may meet at the same vertex ($X$ or
$O$). If the two paths meet at the same vertex, $\delta_{uv} = 0$,
$\delta_{uv} = 1$ otherwise. According to the above classification
of nodes, $\Theta_t^{1,2}$ can be expressed as
\begin{eqnarray}\label{cross05}
  \Theta_t^{1,2} &=& \sum_{i=1}^{7}\sum_{j=1}^{7}
  \Bigg(N_{t-1,P_{j}}\,d_{t-1,P_{i}}+N_{t-1,P_{i}}\,d_{t-1,P_{j}} \nonumber \\
  &\quad& +N_{t-1,P_{i}}\,N_{t-1,P_{j}}\delta_{P_{i}P_{j}}\Bigg),
\end{eqnarray}
where $P_i$ and $P_j$ are node classes of $A_{t-1}^{(1)}$ and
$A_{t-1}^{(2)}$, respectively. The quantity $\delta_{P_{i}P_{j}}$
may equal to either 0 or 1, which are related to two kinds of paths:
paths from nodes in $P_i$ to the three peripheral vertices of
$A_{t-1}^{(1)}$, and paths from nodes in $P_j$ to the three
peripheral vertices of $A_{t-1}^{(2)}$. If these two sorts of paths
may meet at the same peripheral vertex, $\delta_{P_{i}P_{j}}=0$; If
they do not meet, $\delta_{P_{i}P_{j}}=1$. Combining previous
equations and results, we get the final expression for
$\Theta_t^{1,2}$,
\begin{eqnarray}\label{cross06}
\Theta_t^{1,2} &=&\frac{1}{2178} \exp(-\text{i} \pi  t)  \Bigg( 2^t (-62-16\, \text{i} \sqrt{2}) \nonumber \\
&\quad& +11\,\text{i}\, 2^\frac{t}{2} (10\,\text{i}+7 \sqrt{2}) \exp
\left(\frac{i \pi t}{2}\right) \nonumber \\ &\quad&
-11\,\text{i}\,2^\frac{t}{2} (-10\,\text{i}+7 \sqrt{2}) \exp
\left(\frac{3\text{i} \pi
t}{2}\right) \nonumber \\
&\quad& +\text{i}
2^{1+t} (31\,\text{i}+8 \sqrt{2}) \exp(2 \text{i} \pi t) +\exp(\text{i} \pi t) \nonumber \\
&\quad& \left(-484+11\cdot3^{2+t}+9^{3+t}+22\,t\cdot9^{1+t} \right)
\Bigg).
\end{eqnarray}

\begin{figure}
\begin{center}
\includegraphics[width=0.40\textwidth]{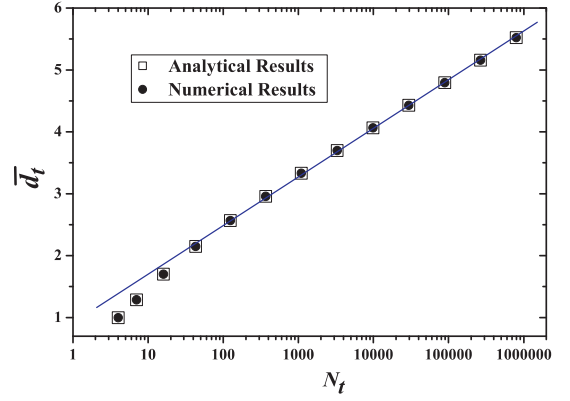} \
\end{center}
\caption[kurzform]{\label{distance} Average path length
$\bar{d}_{t}$ versus network order $N_{t}$ on a semilogarithmic
scale. The solid line is a guide to the eye, which clearly shows
that the APL scales logarithmically with the network size.}
\end{figure}

Inserting Eq.~(\ref{cross06}) into Eq.~(\ref{total03}) and using the
initial condition $D_{1} =6$, we have

\begin{eqnarray}\label{total04}
D_{t} &=&\frac{1}{330 (-3+\text{i} \sqrt{2})
(-3\,\text{i}+\sqrt{2})}\, \Bigg[6655\,\text{i}+155 (-\text{i})^t
2^{\frac{1}{2}+\frac{t}{2}} \nonumber \\
&\quad& -155 \,\text{i}^t 2^{\frac{1}{2}+\frac{t}{2}}+5\,\text{i}
(-\text{i})^t 2^{4+\frac{t}{2}}+5\,\text{i} \,\text{i}^t
2^{4+\frac{t}{2}}\nonumber\\ &\quad& + 31\,\text{i} (-1)^t 2^{2+t}
+133\,\text{i}\, 3^{3+t}+40\,\text{i}\,3^{2+2 t} \nonumber \\
&\quad& + 55 \,\text{i}\,t\,3^{1+t}+55\,\text{i}\,t\, 3^{1+2 t}
\Bigg].
\end{eqnarray}
Then the exactly analytic expression for average path length can be
obtained as
\begin{eqnarray}\label{apl02}
\bar{d}_{t} &=&\frac{4}{1815 (15+8\cdot3^t+9^t)}\Bigg(6655+5
(-\text{i})^t 2^{4+\frac{t}{2}}\nonumber \\ &\quad& +5\,\text{i}^t
2^{4+\frac{t}{2}}-155\,\text{i} (-\text{i})^t 2^{\frac{1+t}{2}}+
155\,\text{i}\,\text{i}^t 2^{\frac{1+t}{2}} \nonumber \\ &\quad& +31
(-1)^t 2^{2+t}+133\cdot3^{3+t}+40\cdot9^{1+t} \nonumber \\ &\quad& +
55\,t \cdot3^{1+t} (1+3^t)\Bigg),
\end{eqnarray}
which can be simplified according to $t$ as follows:
\begin{equation}\label{apl03}
\bar{d}_{t}= \left\{ \begin{array}{cc}
\frac{4[6655 + 5 \cdot 2^{5+\frac{t}2} +31 \cdot 2^{2 + t} +133 \cdot 3^{3 + t} + 40 \cdot 9^{1+t} + 55 \cdot 3^{1+t}(1 + 3^t)t]}{1815(15 +8 \cdot 3^t + 9^t)},\\
\frac{4[6655 - 155 \cdot 2^\frac{3 + t}2 -31 \cdot 2^{2 + t} +133 \cdot 3^{3 + t} + 40 \cdot 9^{1+t} + 55 \cdot 3^{1+t}(1 + 3^t)t]}{1815(15 +8 \cdot 3^t + 9^t)},\\
\frac{4[6655 - 5 \cdot 2^{5+\frac{t}2} +31 \cdot 2^{2 + t} +133 \cdot 3^{3 + t} + 40 \cdot 9^{1+t} + 55 \cdot 3^{1+t}(1 + 3^t)t]}{1815(15 +8 \cdot 3^t + 9^t)},\\
\frac{4[6655 + 155 \cdot 2^\frac{3 + t}2 -31 \cdot 2^{2 + t} +133 \cdot 3^{3 + t} + 40 \cdot 9^{1+t} + 55 \cdot 3^{1+t}(1 + 3^t)t]}{1815(15 +8 \cdot 3^t + 9^t)},\\
\end{array}
\right.
\end{equation}
for $t\equiv 0,1,2,3 \pmod{4}$ are given consecutively. In the
infinite network size (i.e., large $t$), $\bar{d}_{t} \simeq
\frac{4}{11}t\sim \ln N_t$. Thus, the APL grows logarithmically with
increasing size of the network. We have checked our analytic result
against numerical calculations for different network size up to
$t=13$ which corresponds to $N_{13}=797164$. In all the cases we
obtain a complete agreement between our theoretical formula and the
results of numerical investigation, see Fig.~\ref{distance}.

We finally make a comment on the result presented
in~\cite{AnHeAnSi05}. Using a combination of numerical experiments
and heuristic arguments, Ref.~\cite{AnHeAnSi05} advances claims
about the interesting quantity of average path length. The authors
of~\cite{AnHeAnSi05} claimed that $\bar{d}_{t} \propto (\ln
N_t)^{3/4}$. They thus concluded that ANs belong to a new class of
networks which interpolates  between \emph{small} ($\bar{d}_{t}
\propto \ln N_t$) and \emph{ultrasmall} ($\bar{d}_{t} \propto \ln
\ln N_t$) networks~\cite{CoHa03}. While the combinatorial arguments
and simulation results are sound, the behavior of APL as $t$ grows
is quite difficult to estimate numerically. Just as it is hard to
verify with direct computation that the harmonic series diverges,
the experimental results in~\cite{AnHeAnSi05} underestimate the
asymptotic behavior of APL. In contrast to previous claims, our
obtained precise value for APL shows that the conclusion
of~\cite{AnHeAnSi05} seems questionable. The self-similar structure
of Apollonian network allows us to compute precisely the quantity
APL, which is difficult to apprehend through simulation.

In conclusion, in this report we have derived analytically exact
solution for the average path length of Apollonian networks
which has been attracting much research interest. 
We found that in the infinite network size
limit Apollonian networks are small-world, the APL scales
logarithmically with network size. However, some authors have been
proved that conventional random scale-free networks with degree
exponent $\gamma < 3 $ show ultra small-world
property~\cite{CoHa03,ChLu02,DoMeSa03}. In future, it is interesting
to do further studies to reveal this dissimilarity between
Apollonian networks and those stochastic scale-free networks, which
have different scaling for APL. Finally, we believe that the
analytical calculation here can guide and shed light on related
studies for deterministic network models.

 The authors would like to thank T. Zhou and F.
Comellas for many valuable suggestions. This research was supported
by the National Basic Research Program of China under grant No.
2007CB310806, the National Natural Science Foundation of China under
Grant Nos. 60496327, 60573183, 90612007, 60773123, and 60704044, the
Postdoctoral Science Foundation of China under Grant No.
20060400162, the Program for New Century Excellent Talents in
University of China (NCET-06-0376), and the Huawei Foundation of
Science and Technology (YJCB2007031IN).


\begin{references}

\bibitem{CoRoTrVi07}
L. da. F. Costa, F. A. Rodrigues, G. Travieso, and P. R. V. Boas,
Adv. Phys. {\bf 56}, 167 (2007).


\bibitem{AlBa02} R. Albert and A.-L. Barab\'asi,
       Rev. Mod. Phys. {\bf 74}, 47 (2002).

\bibitem{DoMe02} S. N. Dorogvtsev and J.F.F. Mendes,
Adv. Phys. {\bf 51}, 1079 (2002).

\bibitem{Ne03} M. E. J. Newman,
SIAM Review {\bf 45}, 167 (2003).


\bibitem{BoLaMoChHw06}
S. Boccaletti, V. Latora, Y. Moreno, M. Chavezf, and D.-U. Hwanga,
Phy. Rep. {\bf 424}, 175 (2006).

\bibitem{WaSt98} D.J. Watts and H. Strogatz,
        Nature (London) {\bf 393}, 440 (1998).

\bibitem{CoHa03}
R. Cohen and S. Havlin, Phys. Rev. Lett. {\bf 90}, 058701 (2003).

\bibitem{ChLu02}
 F. Chung and L. Lu, Proc. Natl. Acad. Sci. U.S.A. {\bf 99}, 15879 (2002).

\bibitem{DoMeSa03}
S. N. Dorogovtsev, J. F. F. Mendes, and A.N. Samukhin, Nucl. Phys.
{\bf 653}, 307 (2003).

\bibitem{FrFrHo04}
A. Fronczak, P. Fronczak, and J. A. Ho{\l}yst, Phys. Rev. E {\bf
70}, 056110 (2004).

\bibitem{Lolo03}
W. S. Lovejoy, C. H. Loch, Soc. Netw. {\bf 25}, 333 (2003).

\bibitem{HoSiFrFrSu05}
J. A. Ho{\l}yst, J. Sienkiewicz, A. Fronczak, P. Fronczak, and K.
Suchecki, Phys. Rev. E {\bf 72}, 026108 (2005).


\bibitem{DoMeOl06}
S. N. Dorogovtsev, J. F. F. Mendes, and J. G. Oliveira, Phys. Rev. E
{\bf 73}, 056122 (2006).


\bibitem{bo73}
D.W. Boyd, Canadian Journal of Mathematics {\bf 25}, 303 (1973).


\bibitem{AnHeAnSi05} J.S. Andrade Jr., H.J. Herrmann, R.F.S. Andrade and L.R. da Silva,
Phys. Rev. Lett. {\bf 94}, 018702 (2005).


\bibitem{DoMa05} J.P.K. Doye and C.P. Massen.
Phys. Rev. E {\bf 71}, 016128 (2005).


\bibitem{BaRaVi01} A.-L. Barab\'asi, E. Ravasz, and T. Vicsek,
          Physica A  {\bf 299}, 559 (2001).

\bibitem{DoGoMe02} S.N. Dorogovtsev, A.V. Goltsev, and J.F.F. Mendes,
          Phys. Rev. E {\bf 65}, 066122 (2002).

\bibitem{JuKiKa02} S. Jung, S. Kim, and B. Kahng,
        Phys. Rev. E {\bf 65}, 056101 (2002).

\bibitem{CoFeRa04} F. Comellas, G. Fertin and A. Raspaud,
Phys. Rev. E {\bf 69}, 037104 (2004).

\bibitem{ZhRoZh07}
Z. Z. Zhang, L. L. Rong, and S. G. Zhou,
Physica A {\bf 377}, 329 (2007).

\bibitem{RaSoMoOlBa02}
E. Ravasz, A.L. Somera, D. A. Mongru, Z. N. Oltvai, and A.-L.
Barab\'asi, Science {\bf 297}, 1551 (2002).

\bibitem{RaBa03}
E. Ravasz and A.-L. Barab\'asi, Phys. Rev. E {\bf 67}, 026112
(2003).

\bibitem{CoOzPe00}
F. Comellas, J. Oz\'on, and J.G. Peters, Inf. Process. Lett. {\bf
76}, 83 (2000)

\bibitem{ZhRoGo06}
Z.Z. Zhang, L.L Rong and C.H. Guo, Physica A {\bf 363}, 567 (2006).

\bibitem{BeOs79}
A. N. Berker and S. Ostlund, J. Phys. C {\bf 12}, 4961 (1979).

\bibitem{Hi07}
M. Hinczewski, Phys. Rev. E {\bf 75}, 061104 (2007).

\bibitem{ZhZhZo07}
Z.Z. Zhang, S. G. Zhou, and T. Zou, Eur. Phys. J. B {\bf 56}, 259
(2007).

\bibitem{BaFeDa06}
L. Barri\'ere, F. Comellas, and C. Dalf\'o,
J. Phys. A {\bf 39}, 11739 (2006).


\bibitem{ZhZhFaGuZh07} Z.Z. Zhang, S.G. Zhou, L.J. Fang, J.H. Guan,
Y.C. Zhang, EPL {\bf 79}, 38007 (2007).


\bibitem{RoHaAv07} H. D. Rozenfeld, S. Havlin, and D. ben-Avraham, New J. Phys. {\bf 9},
175 (2007).



\bibitem{AnHe05} R.F.S. Andrade and H.J. Herrmann,
Phys. Rev. E {\bf 71}, 056131 (2005).

\bibitem{LiGaHe04} P. G. Lind, J.A.C. Gallas, and H.J. Herrmann,
Phys. Rev. E {\bf 70}, 056207 (2004).


\bibitem{ZhRoZh06}
Z. Z. Zhang, L. L. Rong, and S. G. Zhou,
 Phys. Rev. E, {\bf 74}, 046105 (2006).



\bibitem{Bobe05}
E. Bollt, D. ben-Avraham, New J. Phys. {\bf 7}, 26 (2005).




\bibitem{HiBe06}
M. Hinczewski and A. N. Berker, Phys. Rev. E {\bf 73}, 066126
(2006).


\bibitem{BaAl99} A.-L. Barab\'asi and R. Albert,
       Science {\bf 286}, 509 (1999).


\bibitem{ZhCoFeRo06}
Z.Z. Zhang, F. Comellas, G. Fertin and L.L. Rong,
J. Phys. A {\bf 39}, 1811 (2006).


\bibitem{MaHeRi04}
R. Mahmoodi Baram, H.J. Herrmann, and N.Rivier, Phys. Rev. Lett.
{\bf 92}, 044301 (2004).

\bibitem{ZhYaWa05} T. Zhou, G. Yan, and B.H. Wang,
Phys. Rev. E {\bf 71}, 046141 (2005).


\bibitem{ZhRoCo06}
 Z.Z. Zhang, L.L. Rong and F. Comellas,
Physica A {\bf 364}, 610 (2006).


\bibitem{ZhZh07}
Z. Z. Zhang  and S. G. Zhou,
 Physica A {\bf 380}, 621 (2007).




\end{references}
\end{document}